# Phonon Optimized Potentials

Andrew Rohskopf [1], Hamid R. Seyf,[1] Kiarash Gordiz,[1] Asegun Henry[1,2,3]

**Affiliations:**

[1] George W. Woodruff School of Mechanical Engineering, Georgia Institute of Technology, Atlanta, GA 30332, USA.

[2] Heat Lab, Georgia Institute of Technology, Atlanta, GA 30332, USA.

[3] School of Materials Science and Engineering, Georgia Institute of Technology, Atlanta, GA 30332, USA.

## Abstract

Molecular dynamics (MD) simulations have been extensively used to study phonons and gain insight, but direct comparisons to experimental data are often difficult, due to a lack of empirical interatomic potentials (EIPs) for different systems. As a result, this issue has become a major barrier to realizing the promise associated with advanced atomistic level modeling techniques. Here, we present a general method for specifically optimizing EIPs from *ab initio* inputs for the study of phonon transport properties, thereby resulting in phonon optimized potentials (POPs). The method uses a genetic algorithm (GA) to directly fit to the key properties that determine whether or not the atomic level dynamics and most notably the phonon transport are described properly.

## Introduction

Over the last 25 years, the usage of molecular dynamics (MD) simulations to study phonons has grown markedly. The reason MD is a useful tool for studying phonons is because of three primary advantages/features: (1) it naturally includes anharmonicity to full order, (2) it naturally includes full atomic level details of the structure (i.e. composition, defects, boundaries, etc.), and (3) it can access the necessary time ($10^{-6}$-$10^{2}$ ns) and length scales ($10^{-2}$-$10^{3}$ nm) using today's high performance computing hardware.[1,2] However, given the growing interest and usage of MD to study phonons, there has not been a correspondingly large increase in the number and fidelity of direct comparisons to experimental data. This is largely because of the lack of suitable and accurate empirical interatomic potentials (EIPs) that can describe the various chemistries involved in the actual systems being measured.[3-6] Here it should be emphasized that the EIP is the most important aspect of an MD simulation, because it contains all the physics and in essence, a MD simulation is merely a way of sampling the EIP. Another reason it is difficult to compare MD results directly to experiments is because there is often insufficient detail known about the atomic level structure of the samples being measured to facilitate construction of an accurately representative MD supercell. However, atom probe tomography could serve as an experimental means to bridge this gap. Nonetheless, it is because of these challenges that fair and rigorous comparisons between MD simulation data and experiments are lacking, and this has stifled the ability for theorists to explain and/or predict material properties and anomalous behaviors observed in experiments, which would further our understanding.[6-10]

Given the reality of these barriers to scientific advancement, the development of a means by which one can quickly, easily and accurately create EIPs for the purposes of studying phonon transport has become a grand challenge for the field. Here, the notion that one should be able to quickly and easily parameterize EIPs is important to emphasize, as there have been major advances over the last 25 years that enable the creation of accurate EIPs[11-15]. Most notably, these advances include the proliferation of first principles methods such as density functional theory (DFT), which can be used to generate data that interatomic potentials can be fit to reproduce. Via direct usage of DFT calculations, one can rather consistently and accurately predict the thermal conductivity of pure homogenous crystalline materials.[6,16,17] The problem, however, is that these methods are currently restricted to infinitely large crystals, where periodicity can be exploited to drastically reduce the number of atoms that have to be explicitly included in the calculation. Yet many experiments where interesting and seemingly non-intuitive behaviors occur are outside this realm and require the ability to study much larger numbers of atoms ($10^3$-$10^6$) and time scales that are computationally infeasible with present *ab initio* methods. Thus, there is still a great need and value to having EIPs, which will enable the study of disordered materials such as alloys and amorphous materials as well as nano/microstructures.

It is also critical to appreciate that having the ability to generate an EIP that can correctly describe the vibrations of atoms would be extremely powerful for applications of MD beyond the realm of phonon transport. For example, such potentials can describe the background atoms in MD simulations of reactions,[18] crack propagation,[19] or diffusion by employing methods such as learn on the fly technique[20] or other similar strategies.[21] This can greatly reduce the computational expense of studying such problems. Similarly, EIPs can be co-optimized to accurately capture defect energies along with the vibrational characteristics, which would be extremely useful in extrapolating the results of DFT towards the calculation of thermodynamics properties in alloyed systems as well as vibrational entropy.[22,23] The major problem, however, is simply that the time and effort required to parameterize an accurate EIP is so great that it often serves as a deterrent for the actual problem of interest. The reason, the time for creation is often so long, is because one must first identify a suitable functional form for the EIP and then perform a highly multi-dimensional optimization to find the best set of N parameters for the EIP that describe the system and properties of interest.

In the past, many efforts to produce such EIPs involved modification of the functional form itself and the properties used for fitting often included experimental measurements of the lattice parameters, elastic constants, phonon frequencies, and other measureable quantities.[24,25] More recently, efforts to fit EIP parameters have shifted to *ab initio* data, which offers a more direct connection to atomistic level quantities such as forces, energies and stresses on supercells, which cannot be easily determined experimentally[14,26]. However, the N-dimensional optimization problem to find the best EIP parameters is still daunting, and in the past has been partially guided by chemical/physical intuition into the system of interest. For this reason, many popular functional forms for EIPs have been modified for different systems to achieve improved accuracy.[27,28] Consequently, many MD investigations resort to using whatever parameters can be found somewhere in the literature for an EIP that has already been coded, and have been applied to the specific system of interest. Furthermore, this usage of standard EIPs and parameters from literature often happens regardless of whether the EIP accurately describes the phonon transport or not, simply due to a lack of options. Thus, the grand challenge has been to develop a quick and easy method for creating EIPs that accurately describe phonons. The emphasis here on the process being quick and easy, is so that the EIPs can be created and employed with minimal effort and the major intellectual investment can remain focused on the MD and phonon transport instead of the prerequisite issue of finding a suitable EIP. Herein we describe a methodology and a set of freely available codes that implement the methodology that can overcome this barrier of generating what we have termed phonon optimized potentials (POPs).

**Fundamental Tenets and Goals of the POPs Approach**

Before explaining the methodology itself it is useful to first highlight its goals – the basic tenets that underpin it and several questions we seek to answer with the first example usage of it, which is described later:

1) Many EIP functional forms are typically overdesigned for the study of phonons, so we hypothesize that many solutions exist to describe such properties. Popular EIPs are designed to describe various configurations of the atoms, and most notably different atomic coordinations[29-31]. However, it is most often the case that when one seeks to study phonons, all atoms by definition vibrate around their equilibrium sites[32] and the atomic coordination and configuration are fixed for all atoms throughout the entire MD simulation[33]. Given this great overdesign of standard EIPs to describe unnecessary regions of phase space for the study of phonons, we hypothesize that many parameter sets exist for such EIPs that describe phonons, but may not describe other properties. Nonetheless, since the focus herein is to optimize for phonons, such EIPs would still be considered accurate for the purposes herein, despite their limited transferability. It is important to clarify here that transferability is not a goal/objective of the POPs framework, since it is not essential to describing phonon transport, which is most often focused on a fixed set of equilibrium positions.

2) In order for an EIP to be optimized for describing phonons, it is hypothesized that the key quantities that must be well described are the total energy and its derivatives. Taking results from the fluctuation-dissipation theorem[34,35] as a basis for describing phonon transport properties such as thermal conductivity[33,35] and interface conductance[36], this hypothesis is based on the idea that one will obtain the correct transport properties if all of the forces (including correct individual force components[37]) and velocities of the atoms are correct. Also, from the formalism developed for crystalline thermal conductivity[37] it is known that if one can correctly compute all the derivatives of the energy with respect to atomic displacements, one should theoretically properly describe phonon-phonon interactions[16].

Therefore it is hypothesized that an EIP is optimized for phonons when it accurately reproduces the derivatives of the potential energy with respect to the atomic displacements. Although, in concept one would need an infinite number of derivatives to be exact, we note that in practice only the energy and its first three or four derivatives ($E$, $\frac{dE}{dr_i}$, $\frac{d^2E}{dr_i dr_j}$, $\frac{d^3E}{dr_i dr_j dr_k}$ ) are actually needed for most systems/temperatures, but higher order terms can be included as deemed necessary. Here, it is also important to clarify that the goal of POPs is to make EIPs that replicate the results of *ab initio* calculations and not necessarily experiments. In this sense the goal is to make POPs based purely on first principles data, thereby enabling them with predictive power. If there are discrepancies between the *ab initio* data and experiments, then an improved *ab initio* description may be required.

3) Assuming the preceding hypothesis is correct, then one can set as components of an objective function, the relative error in energy and its derivatives to assess the viability of a potential to describe phonon properties. This then provides a universal scale upon which any EIP can be assessed. For example, one can assess that a given EIP reproduces the energy of the *ab initio* model within 3%, the forces within 5%, the second derivatives within 10% on average, and so on. In this respect one can also invoke statistical techniques as well (i.e., standard deviation, root mean squared etc.) to better assess the EIP accuracy. It may also happen that a certain functional form is unable to reduce the error in energy and its derivatives to say below ~ 40%. Such a functional form may therefore not be suitable for the system of interest and it is important that the methodology be able to determine this. It is therefore a goal of the methodology to enable assessment of the functional form itself. This requires that somehow the search for parameters be exhaustive/complete enough to allow one to conclude that a particular functional form is simply incapable of describing the system within the target accuracy.

4) A major goal of the POPs approach is to make the creation of EIPs for a given system easy and quick. Here, the term "easy" is to imply that minimal input, chemical insight and management on the part of the user is required and instead the procedure itself is capable handling most of the effort. Furthermore, it is highly desirable to have the approach be built in such a way that it does not require new coding on the part of the user, which inherently requires time for debugging and is a strong deterrent. In this respect, the goal was to streamline the process and reduce the time required to obtain a potential to the order of one week, assuming the *ab initio* calculations are already available.

Considering the aforementioned goals/tenets and hypotheses, several questions arise that we seek to answer in herein, namely: (1) Conceptually, the total forces on the atoms are all that is needed to uniquely determine the trajectory, but it is not clear if an EIP is solely designed to reproduce the forces, whether it will correspondingly be optimized for phonons by default. Thus, it is not clear how important the energy and its derivatives, as well as other properties such as stresses are in the creation of a POP. Thus, the question arises, is it possible for an EIP to exhibit accurate forces, but somehow exhibit other problems, i.e., become unstable when

MD simulations are executed? (2) Considering the aforementioned hypothesis in Tenet 1 regarding the possibility of finding multiple/many sets of parameters that nearly degenerately minimize the target objective function, it is not clear if all of such solutions will exhibit commonalities, e.g., they can somehow be reduced to a single unique best solution, or if some are uniquely different, exhibiting drastically different parameters that somehow still yield similar objective function values. (3) It is not clear *a priori* according to Tenet 3 whether common and standard EIP functional forms will be able to accurately reproduce *ab initio* results at all. Thus, herein we seek to determine if the proposed approach can actually yield useful answers that can at least reproduce thermal conductivity within ~ 10%.

**POPs Methodology**

The POPs methodology described in the following has been implemented in a freely available C++ code[38] and can be run massively in parallel. The approach uses a genetic algorithm (GA) to search for parameters that minimize an objective function representing some form of error between candidate EIPs and *ab initio* results. The GA is a metaheuristic that mimics natural selection to guide the algorithm towards a solution of a multi-dimensional problem.[39-41] Gradient based methods work well for problems with less dimensions and much less non-linearity. However, for more complex problems with large numbers of dimensions and strongly non-linear behavior, gradient based methods easily fail and alternative schemes such as a GA are needed.[12,42] Since the GA approach itself inherently searches for random perturbations to potential solutions (i.e., via crossover and mutation), one is in practice guaranteed to exhaustively exploit many local minima if many trials are run in parallel. Many parallel trials also exhausts all possibilities for a given functional form, which allows one to determine its suitability for a given system (Tenet 3). For example, if many GA trials yield a high percentage of undesirable solutions, it can be said that the functional form may not be an appropriate candidate for the system of interest. Additional details associated with the POPs GA are described in the SI.

The code couples with the open source MD software LAMMPS[43] as a calculator for EIPs, and the open source code Alamode[44] as a calculator for interatomic force constants (IFCs). LAMMPS was selected because it has many standard EIPs already coded within it as well as many common variants. This prevents users from having to write new code to try different EIP functional forms, which contributes strongly towards making the process "easy" (Tenet 4). Alamode was selected because it can take as inputs a series of arbitrary atomic displacements and forces to then compute $2^{nd}$, $3^{rd}$ and $4^{th}$ order derivatives of the energy with respect to the atomic displacements, as required by Tenets 2 and 3 to create a POP.[44] Using LAMMPS and Alamode as calculators for EIP properties, Tenet 3 requires that an objective function is needed to determine the viability of an EIP compared to reference values. In general, the objective function consists of different quantities that are weighted according to their relative importance on a normalized scale via Eq. 1:

$$Z = \sum_i w_i z_i \qquad (1)$$

where each $z_i$ generically represents a normalized error between the values produced by a candidate EIP as compared to the *ab initio* values, weighted by a factor $w_i$. In this work, Eq. 1 is a sum of the weighted errors of Hellman-Feynman forces, energies, stresses, and 2$^{nd}$ and 3$^{rd}$ order IFCs with weights $w_f$, $w_e$, $w_s$, $w_{ifc-2}$, and $w_{ifc-3}$, respectively. These quantities are inspired from the discussion in Tenets 2 and 3, and the stresses were found to be necessary to ensure crystal stability in MD simulations. The format of the error $z_i$ for forces, energies, stresses, and IFCs will be described elsewhere.

Lastly the interface with *ab initio* results has been generalized so that users can employ any code of interest. The inputs are simply a series of supercell snapshots containing atomic coordinates, the total energy, the individual atom total forces and supercell stresses for configurations with atoms randomly displaced from equilibrium positions as well as different volumes. Several previous works have indicated that using random displacements, or even displacements from *ab initio* MD trajectories, is highly effective in capturing anharmonicity.[44-46] Each randomly displaced configuration contains information about many interatomic interactions and not just a subset of atoms as is the case for the direct displacement method.[16] As a result, it has been found that using random displacements reduces the search space by providing more valuable information on anharmonicity and other interactions that result in stability during a MD simulation.

**Results and Discussion: The Example Cases of Silicon**

With the preceding framework implemented we then tested the code on the two most popular example systems that are of very high technological/applications interest, namely crystalline silicon (c-Si). The specific case of creating EIPs for c-Si is discussed here in detail, because it highlights important features and answers to the questions/hypotheses outlined in the Tenets. Through a process that largely consisted of trial and error, we have converged to a generally suitable weights for the objective function and an important test for stability during an actual MD simulation. It is believed that these settings are rather general and will translate to other systems, but this will require much more extensive verification. Originally it was found that purely fitting to forces often resulted in unstable dynamics, thus answering question (1). In this respect we have found that there are three important necessary, but individually insufficient quantities that must be accurately reproduced in order to consistently obtain stable MD simulations, namely forces, energies, stresses within ~ 10% of the *ab initio* results. More specifically it is important to: (1) Fit the energy-volume curve shape to ensure the correct lattice parameter and supercell dimensions minimize the energy. (2) It is important to fit to the stresses so that cell dimensions are preserved dynamically once atoms move away from their equilibrium sites. (3) It is, of course, important to fit the forces, since they determine the trajectory and ensure atoms vibrate properly around their equilibrium sites with correctly oriented and scaled restoring forces/accelerations. In general, it has been found that if a functional form is able to reproduce all three of these quantities simultaneously with less than ~10% error, the resulting MD will be stable. More specifically, the low error in forces must include relatively large displacements so that the anharmonicity is significantly sampled and it also affects the stability at higher temperatures. With the issue of stability in MD simulations established, we now discuss results for c-Si using three different functional forms.

For many semiconductors it is well known that long range interactions are important,[47] yet the most popular EIPs such as Tersoff,[48] Stillinger-Weber (SW),[49] the environment dependent interatomic potential (EDIP)[29] and others are restricted to first nearest neighbors. Thus, to illustrate the power of the POPs methodology and to describe c-Si more accurately we fit the *ab initio* data with a combination of short ranged and long ranged functional forms. We selected three popular short ranged EIPs: the Stillinger-Weber, Tersoff, and the [50] potential[51] as implemented in LAMMPS and we also added the Buckingham potential and damped-shifted force (DSF) Coulomb potential[50] to Tersoff and SW, and we added the Born-Mayer-Huggins[51] and DSF Coulomb EIPs to Morse. Since the Morse potential is two-body in nature, we added a harmonic three body term of the form $E = K(\theta - \theta_0)^2$ to account for the covalent three body interactions in c-Si. This then yielded three candidate functional forms: Tersoff + Buckingham + Coulomb (TBC), Stillinger-Weber + Buckingham + Coulomb (SWBC), and Morse + 3 Body + Born + Coulomb (M3BC) – all of which are already coded in LAMMPS. The ranges used for each parameter search and a general rationale for selecting the range is described in the SI.

These potentials were fit to snapshots of DFT calculated configurations of 64 atoms in a supercell using the POPs code, whereby it calls LAMMPS as a library to evaluate the energy and forces of candidate parameter sets, along with Alamode to determine the 2$^{nd}$ and 3$^{rd}$ order IFCs. The objective function in Eq. 1 was then minimized with the following relative weightings: $w_f = 0.15$, $w_e = 0.25$, $w_s = 0.2$, $w_{ifc-2} = 0.2$, and $w_{ifc-3} = 0.2$. One of the most interesting and important outputs of the fitting procedure, which was executed in less than 1 day using 4 processors per trial, was that the various fits yielded >50 uniquely different parameter sets that all had less than 10% error in forces, energies, and stresses (Tenet 1). Figure 1 shows the objective function convergence with generations for all 50 trials in three different EIPs. As shown in Figure 1, the SWBC potential was unable to minimize the objective function as well as TBC and M3BC (Tenet 3).

These parameter sets were then tested against 50 configurations that were randomly displaced by a maximum of 0.5 Å, for which they had not been fit to, and their error in forces energies and stresses was < 10%. Thus, they can be substituted for direct DFT calculations to yield the same forces energies and stresses for an arbitrary configuration to within 10%. Here, the importance of the nominal value of ~10% error is illustrated in Fig. 2, as one can visually see the difference between predicted and fitted forces when the error is 50%, versus when it is as low as 10% and 3%. As a point of comparison, consider the differences in forces for the same configurations using different pseudopotentials, which subsequently result in ~15% standard deviation.[52] Thus, errors much greater than 10% are quite substantial and are much larger than the differences one would expect to observe from improving the *ab initio* calculations. Consequently, even though these popular EIPs have been extensively used to model c-Si, the dynamics are not representative of the real materials. Instead, what has been modeled is likely a fictitious material that happens to have similar thermal conductivity, but not the same phonon trajectories as Si and Ge. For perspective, it is important to realize that the most common parameter sets for the Tersoff[48] and SW[49] potentials to describe c-Si result in 35% and 210% error in forces, respectively.

Clearly the 50% level of error in Fig. 2 would lead to incorrect dynamics and would result in unacceptably large errors in phonon transport properties. However, one could make the argument that < 10% error in forces, might directly translate to ~10% error in the heat flux operator[37] and < 10% error in thermal conductivity[33] and interface conductance[36]. The validity of an approximately 10% error in force, energy and stress metric, along with < 10% errors in IFCs is confirmed by estimations of the thermal conductivity for c-Si using the Boltzmann Transport Equation (BTE) within the relaxation time approximation (RTA) as implemented in Alamode.[44] Figures 3 and 4 show the excellent agreement with the DFT derived thermal conductivity that is obtained when fitting to energies, forces, stresses, and $2^{nd}$ and $3^{rd}$ order IFCs employing the TBC (< 5% thermal conductivity error) and M3BC (<15% thermal conductivity error) potentials for c-Si. It is interesting to note here that the SWBC potential was unable to simultaneously minimize all parts of the objective function to within 10% error of DFT, as seen in the poor fit displayed in Figure 1. The SWBC resulted in unstable dynamics and/or much larger (> 50% error in IFCs) disagreement with phonon transport properties, such as dispersion or forces.

These results answer question (2), namely whether or not multiple parameter sets can be determined that can accurately reproduce the trajectories that would have been obtained if a DFT MD calculation could be evaluated at the requisite length and time scales. We have also answered the question as to whether or not some common functional forms (SWBC) simply fail to simultaneously describe all properties in the objective function, as shown in Fig. 1. The fact that the GA approach yields so many candidate solutions suggests that a paradigm shift is needed in the context of POPs, for how parameter sets are even communicated in subsequent literature. In previous studies EIP parameter sets have been typically communicated in tables, because there is usually only one set of best parameters being discussed. Here, however, there are a myriad of parameter sets that appear to all almost equally satisfy the objective function. Thus, it would be cumbersome and exhaustive to try and accurately transcribe the parameters from the POPs output into a manuscript without errors, or force users to do the reverse. Instead, we have simply numbered the solutions according to a prefix corresponding to the functional forms used (i.e., TBC-1, TBC-2, TBC-3, etc.) and included sample input files for LAMMPS, which can simply be downloaded and used directly.

For an EIP to be optimized for phonons it is important to fit to the IFCs. The $2^{nd}$ order IFCs are most important and whenever possible $3^{rd}$ order IFCs further improve the description although they are less critical if the total forces are well reproduced. As discussed earlier, it is not only important that an EIP properly reproduces the total force, but more specifically the individual force contributions from different atoms. Inspection of Hardy's flux operator[37] also indicates that it is the respective force contributions, subsequently multiplied by the appropriate velocities that would yield the correct heat flux. The $2^{nd}$ order IFCs are therefore more important to include since the harmonic components of forces usually comprise the overwhelming majority (> 90%) of the forces and energy[53]. This information is not explicitly contained in the total forces and therefore it is important to separately include the $2^{nd}$ order derivatives (e.g., the Hessian matrix) as part of the objective function, since it is possible to have different EIPs reproduce the total forces correctly with drastically different force components.

Phonon dispersion relations for various POPs are shown in Figs 5-6 for TBC and M3BC c-Si. It is also important to include the $2^{nd}$ order IFCs in order to ensure the interactions between atoms are properly scaled and yield the correct dispersion. Results for thermal conductivity can be significantly improved by fitting directly to the $3^{rd}$ order IFCs from DFT as well, since it is possible to have agreement with dispersion but inaccurate thermal conductivity. The main error in phonon frequencies with the TBC and M3BC potentials is seen with their failure to reproduce the flattening of frequencies at the zone boundaries in Figs 5-6, due to the inability of the pair potential to reproduce long range IFCs. This issue, however, has been overcome with other analytical functional forms such as the bond charge model[54] and valence force field model[55,56]. Future work will involve the incorporation of these models into POPs, which are more suitable for describing phonons but not yet currently implemented in the LAMMPS package.

**Conclusion**

We have developed a framework that consistently produces reliable POPs out of any suitable EIP or combination of EIPs from first principles inputs, by employing a GA to find parameters. In our first demonstration of the methodology we have answered three important rather fundamental questions regarding EIP fitting: (1) It is important to fit to forces, energies and stresses for stability in a MD simulation and to describe phonons properly it is important to also fit to $2^{nd}$ order IFCs and if possible $3^{rd}$ order IFCs as well. (2) It was confirmed that common EIPs are overdesigned for the purposes of exclusively modeling phonon transport and thus many nearly degenerate solutions exist that have drastically different parameters. This finding is particularly important, because different solutions could be more transferrable or better/worse at describing other properties/phenomena of interest and thus it is useful to report all such solutions, so that other users can determine the extent to which POPs parameterizations can be used for other properties. (3) We have confirmed that common functional forms for c-Si can in fact reproduce DFT results within ~ 10%, and therefore serve as useful substitutes to enable probing of larger length and time scales accessible to DFT directly. Furthermore, the GA approach proved useful at performing sufficiently exhaustive searches through parameters that one can evaluate the suitability of the functional form itself for a given system.

Future work will involve using POPs for alloys and disordered systems in order to study thermal transport, since the asymmetry of these systems render them computationally intractable to be studied directly by current *ab initio* methods.[16] The creation of POPs will also allow for the study of thermal transport using MD, with much greater fidelity, enabling direct comparisons with experiments, when some of the most general methodologies are employed.[33,36] Also of critical importance is the fact that the POPs methodology can be used for fields outside of thermal transport. For example, functional forms such as REAXFF[57] can be used in the context of the study of chemical reaction kinetics, and common functional forms used herein can be optimized using the same algorithm to study defects, grain boundaries, interfaces and surfaces by first fitting to the most closely accessible DFT configurations. In this way, it is anticipated that the POPs methodology can serve as a significant advancement in many other areas of science/engineering, beyond that of thermal transport, since phonons are important for many other non-heat transfer centered phenomena as well.[58]

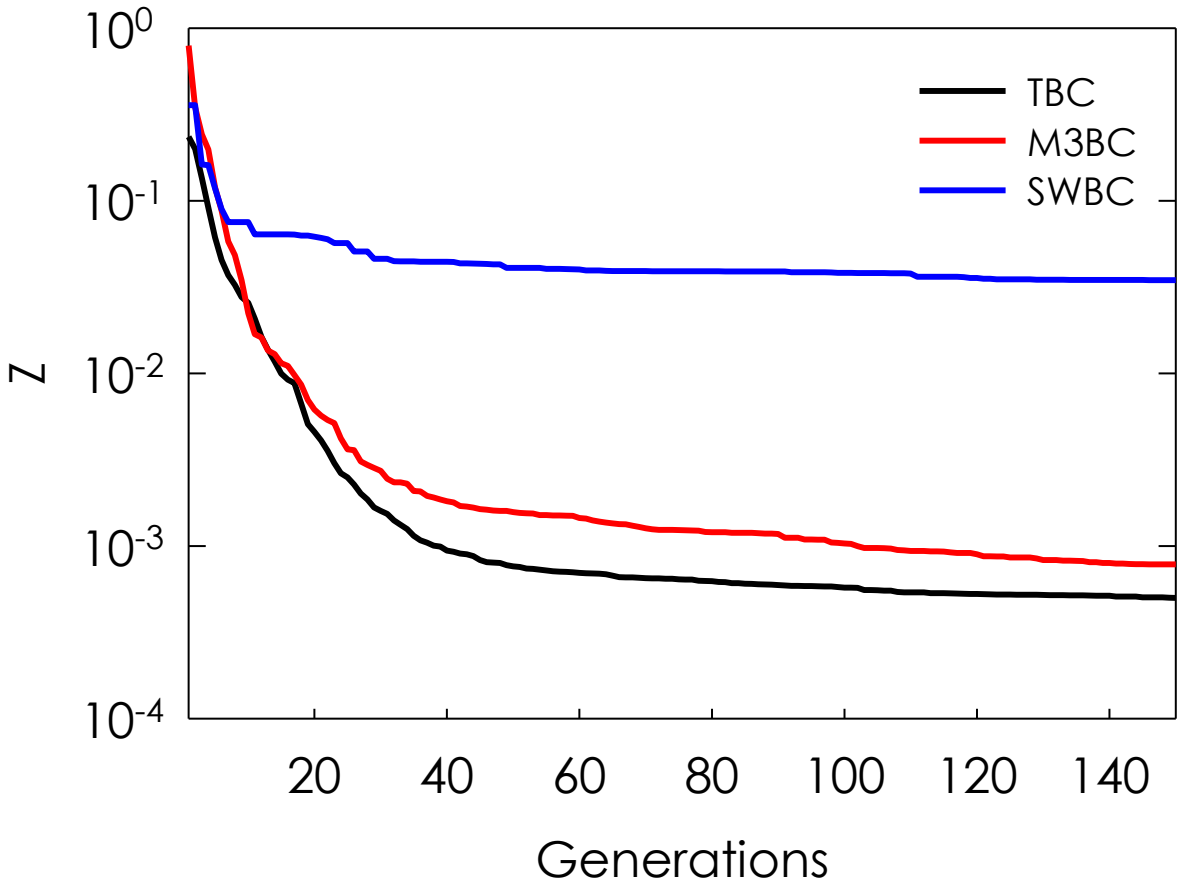

**Figure 1 | Average Z value convergence with generations.** Each line in this figure represents the average Z value (Equations 1) per generation, across all 100 trials performed on the TBC, M3BC, and SWBC potentials. The following weights were used for each Z component: $w_f = 0.15$, $w_e = 0.25$, $w_s = 0.2$, $w_{ifc-2} = 0.2$, and $w_{ifc-3} = 0.2$. This type of analysis shows that some functional forms are able to simultaneously reproduce certain properties better than others. The SWBC, which one average only experienced an order of magnitude decrease from a random guess in Z, could not reproduce all quantities at once. The TBC and M3BC potentials were able to decrease Z by 3 orders of magnitude from a random guess. The small final discrepancy between TBC and M3BC is due to the fact that TBC better reproduced the 3rd order IFCs, which also led to better thermal conductivity agreement as seen with Figures 2 and 3.

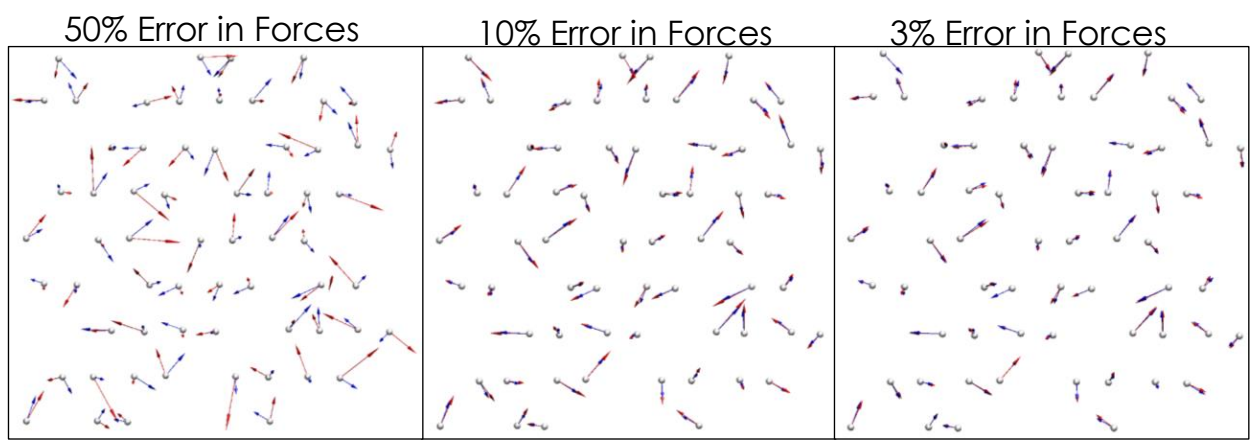

**Figure 2 | Visual representation of different errors in forces.** DFT forces (red vectors) are shown on lattice sites and compared to EIP forces (blue vectors) for 50%, 10% and 3% errors in forces from left to right.

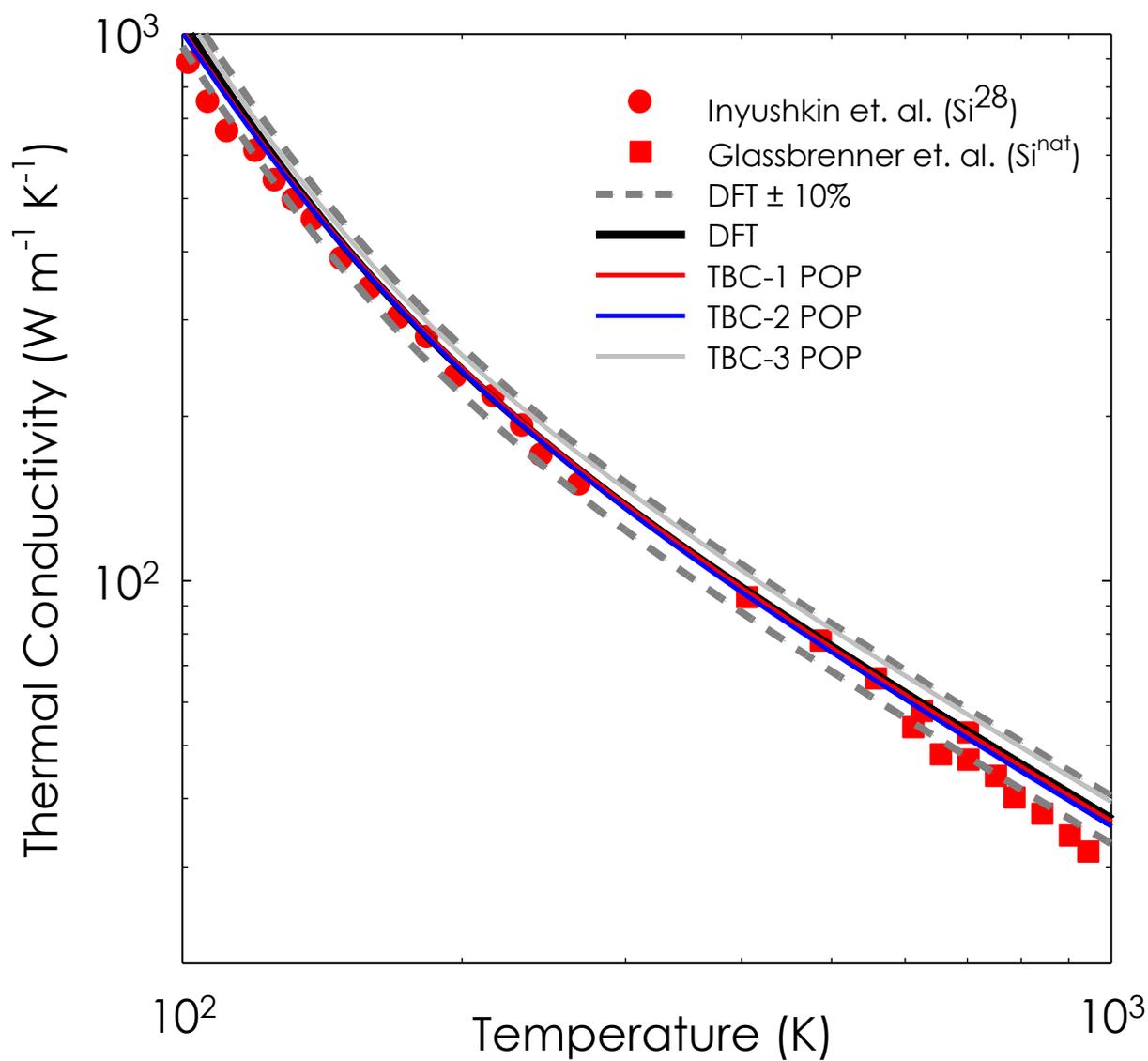

**Figure 3 | TBC, DFT, and experimental[59,60] thermal conductivity vs. temperature for Si.**

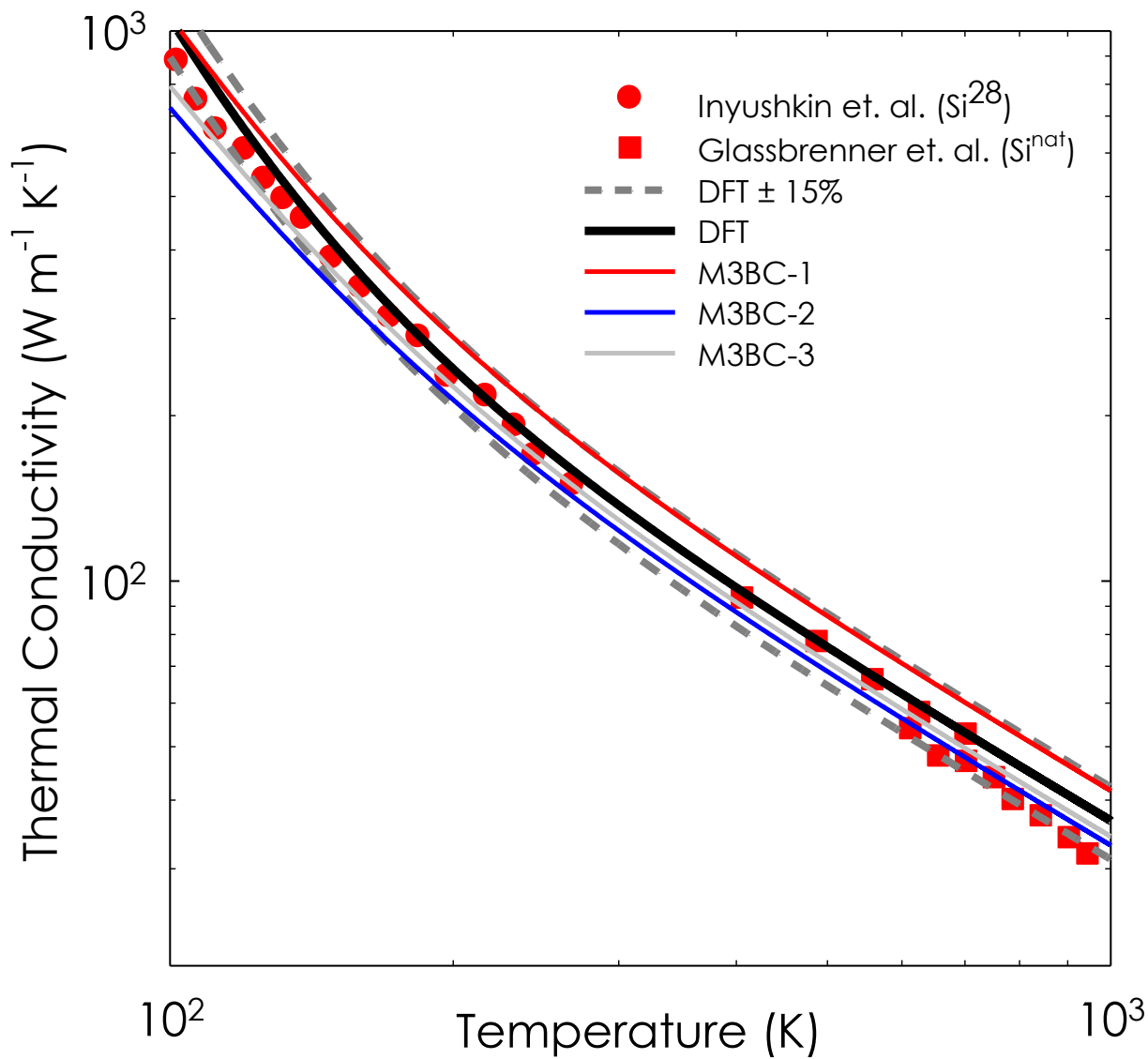

**Figure 4 | M3BC, DFT, and experimental[59,60] thermal conductivity vs. temperature for Si.** While M3BC-3 is within 10% of the DFT thermal conductivity, other M3BC potentials were not able to reach this mark. DFT thermal conductivity ± 15% is therefore displayed to show the performance of the M3BC potential.

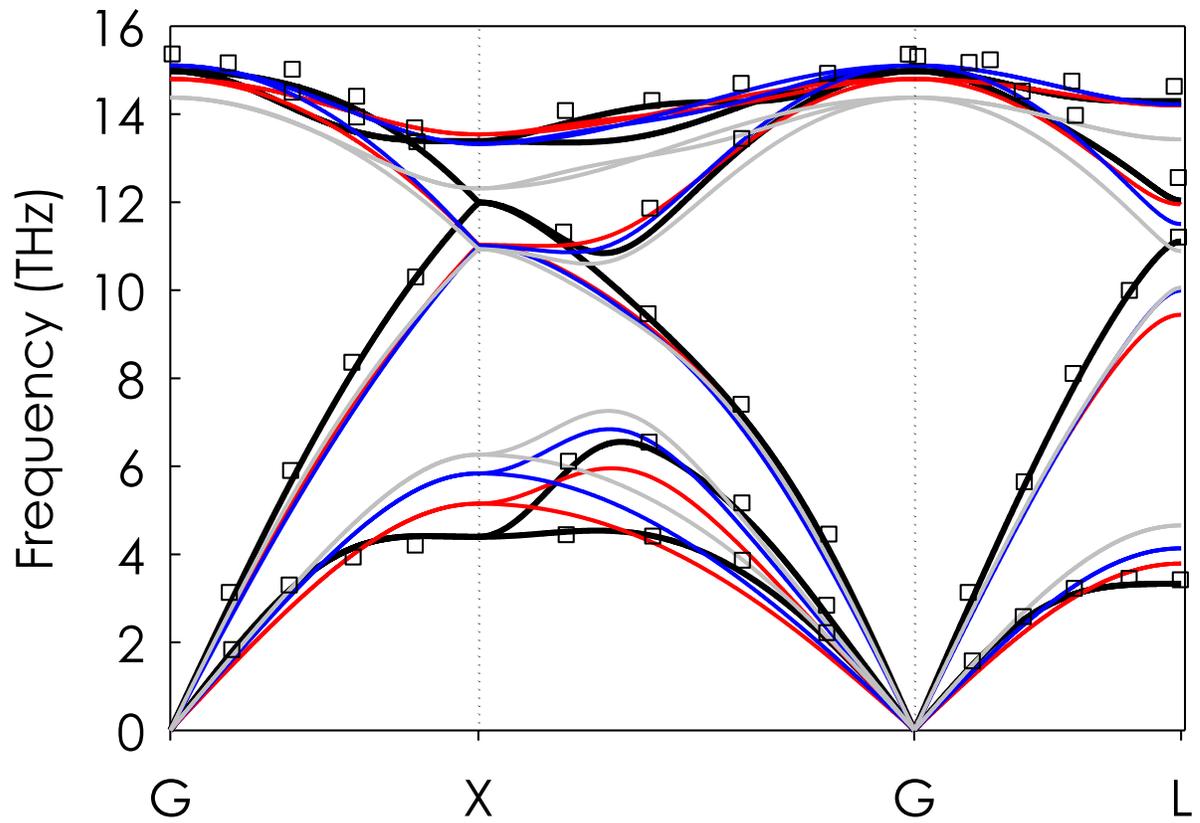

**Figure 5 | TBC, DFT, and experimental[54] phonon dispersion for Si.** DFT (black), TBC-1 (red), TBC-2 (blue), TBC-3 (grey) and experiments (squares) are shown.

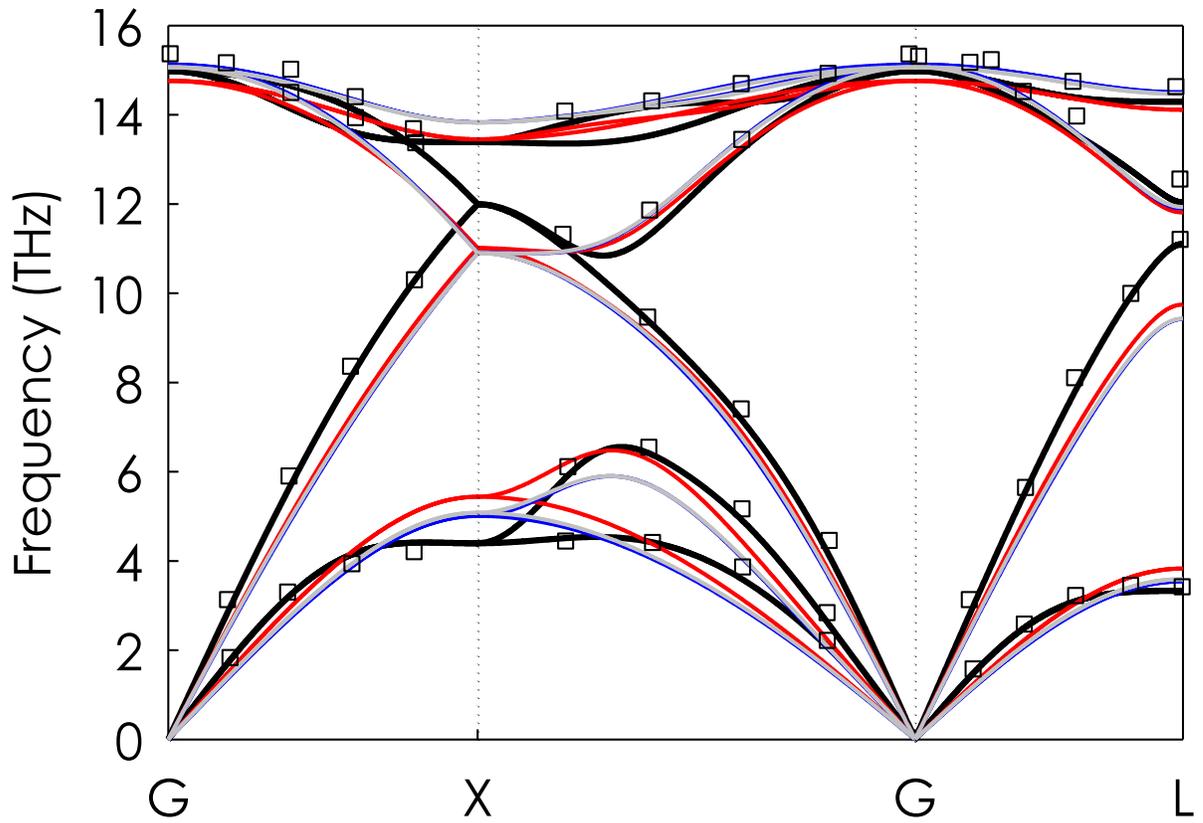

**Figure 6 | M3BC, DFT, and experimental[54] phonon dispersion for Si.** DFT (black), M3BC-1 (red), M3BC-2 (blue), M3BC-3 (grey) and experiments (squares) are shown.